\documentstyle[aaspp4]{article}
\begin{document}

\title {HI 21cm absorption in two low redshift damped Ly$\alpha$
  systems} 

\author{W. Lane, A. Smette\altaffilmark{1}, F. Briggs}
\affil{Kapteyn Astronomical Institute, P.O. Box 800, NL-9700 AV
  Groningen, The Netherlands}
\author{S. Rao, D. Turnshek}
\affil{Dept. of Physics and Astron., Univ.  of Pittsburgh,
  Pittsburgh, PA 15260, USA}
\and 
\author{G.  Meylan}
\affil{ESO, Karl-Schwarzschild-Strasse 2, D-85748 Garching bei
  M\"unchen, Germany}

\altaffiltext{1}{current address: NASA/GSFC, Code 681, Greenbelt, MD 20771,
  USA}

\begin{abstract}  

  We report the discovery of two low redshift HI 21cm absorbers,
  one at $z = 0.2212$ towards the $z_{em} = 0.630$ quasar OI 363
  (B0738+313), and the other at $z = 0.3127$ towards PKS B1127-145
  ($z_{em} = 1.187$).  Both were found during a survey of MgII
  selected systems at redshifts $0.2 < z < 1$ using the new UHF-high
  system at the Westerbork Synthesis Radio Telescope (WSRT).  New
  HST/FOS observations also identify both systems as damped Ly$\alpha$
  (DLa) absorbers.  By comparing the column density from the DLa line
  with that from the HI 21cm line, we calculate the spin temperature,
  T$_s$ of the two systems.  We find $T_s \approx 1000$ K for both of
  these low redshift absorbers.

  For the $z = 0.3127$ system towards PKS B1127-145, two galaxies have
  been previously identified with emission lines at the absorber
  redshift (Bergeron \& Boiss\'e, 1991), with the galaxy at a closer
  projected distance to the quasar assumed to be responsible for the
  absorption system.  An ESO-NTT/EFOSC2 spectrum of a 3rd, fainter
  companion at 3.9 arcsec or 11 $h^{-1}_{100}kpc$ from the line of
  sight of PKS 1127-145 reveals [OIII]4958 and 5007 at $z = 0.3121 \pm
  0.0003$.  We consider this object the most likely to be responsible
  for the 21cm absorption, as it is much closer to the QSO sightline
  than the two galaxies identified by Bergeron \& Boiss\'e.

\end {abstract}

\keywords{quasars: absorption lines --- quasars: individual(PKS
  B1127-145, OI 363) --- galaxies:ISM}

\section{Introduction}

The study of low redshift examples of the quasar absorption line
systems responsible for the damped Ly$\alpha$ (DLa) and HI 21cm
absorption lines is important to help bridge our understanding of
neutral gas-rich systems between those at redshift $z = 0$ and those
at high $z$.  Our knowledge of the neutral gas in nearby spiral
galaxies is mainly based on observations of the HI 21cm line in
emission.  At high redshift, however, we observe the HI 21cm line in
absorption, which can only be seen along a limited number of lines of
sight through the intervening absorber, making detailed knowledge of
the gas characteristics difficult.  The low redshift ($z < 1$) neutral
absorbers are still close enough that both optical and radio data of
reasonable quality can be obtained in order to understand their
kinematics and physical gas characteristics.  Such information is
necessary to build a framework for a correct interpretation of the
higher z counterparts to these systems.

Searches for redshifted HI 21cm absorption can be time-consuming since
radio spectrometers typically observe only relatively narrow
instantaneous bandwidths and only the highest column density QSO
absorption line systems have measurable optical depths in the 21cm
line.  Since DLa systems have high column densities of neutral HI,
they are the most likely objects to have HI 21cm absorption.
Unfortunately for low redshift work, the Ly$\alpha$ line is not
shifted into the optical window until $z \simeq 1.65$, so finding
these lines requires UV spectra to be taken with space telescopes.
This, combined with the small cross section for DLa absorption, means
that only a small number of DLa systems have been identified at low
redshift.  Therefore, alternative selection criteria which are
reasonably effective at finding an HI 21cm absorber must be used.

All known DLa and HI 21cm absorbers have associated low-ionization
metal lines (cf. Lu and Wolfe, 1994) such as the MgII
$\lambda\lambda$2796, 2803 doublet, which can be observed easily in
ground-based spectra down to about $z = 0.1$.  A study of MgII
selected systems using previously existing UV data yielded about 1 DLa
system per 10 MgII systems observed (Rao, et al., 1995).  A similar
statistic exists for HI 21cm absorption in MgII systems (Briggs \&
Wolfe, 1983).  This suggests that MgII can be used to optically select
systems likely to have high column densities of neutral gas observed
either as DLa or HI 21cm absorption.

We are conducting a survey for HI 21cm absorption in low redshift MgII
selected systems using the Westerbork Synthesis Radio Telescope
(WSRT).  In this paper we present two new HI 21cm absorbers from our
survey, one towards PKS B1127-145 at $z = 0.3127$ and the other towards
OI 363 at $z = 0.2212$.  Recent HST/FOS spectra have identified DLa
absorption in both of these objects as well.

\section{Radio Observations}

PKS 1127-145 was observed for 60 minutes on 16 March 1997, and OI 363
for 90 minutes on 27 April 1997 using the new UHF-high system
(700-1200 MHz) at the WSRT.  Observation length was determined by the
amount of time needed to reduce the noise level in the spectra to
$1\%$ or less of the background quasar flux level.  A bandwidth of 2.5
MHz was centered at the expected absorption frequency based on the
MgII absorption redshift of each system.  This was divided into 128
channels at spacings of 5 km s$^{-1}$ for OI 363 and 5.4 km s$^{-1}$
for PKS B1127-145.  We used a uniform spectral taper, so the effective
velocity resolution is 1.2 times the channel spacing, or 6 and 6.5 km
s$^{-1}$ for the respective absorption systems.  Due to limitations in
the DXB spectrometer and the ongoing upgrade at the WSRT, not all of
the interferometers in the array were available.  The OI 363
observation had 24 interferometers in each of two orthogonal linear
polarizations.  For PKS B1127-145, 28 interferometers in one
polarization and 21 in the other were used.

Both data sets were processed identically.  Standard programs in the
NEWSTAR data reduction package were used for initial data calibration
and editing.  The bandpass was calibrated by observations of 3C286
made before and after each object observation.  The data were then
moved to AIPS for further processing.  Using the routine CALIB, the
data were self-calibrated in phase to a point source model of the
field to correct for residual phase errors.  Side-lobe interference
from other strong sources in the field away from the phase center was
not a problem in either data set.  The continuum flux was subtracted
from the visibility data using the algorithm UVLIN, which finds a
linear fit to the complex spectrum measured at each u-v point, and
then subtracts the fit from the spectrum (Cornwell, et al., 1992).

Spectral line image cubes were synthesized from the continuum
subtracted data using the task IMAGR.  The data were given a natural
weighting to reduce the noise as much as possible.  The slight loss of
angular resolution in the WSRT synthesized beam resulting from this
weighting does not affect our spectra, because PKS B1127-145 and OI
363 are both compact sources, much smaller than the beam of the WSRT
at these frequencies.  The image cubes were hanning smoothed in
frequency but not resampled.  The final spectra are shown in figs. 1
and 2.  Absorption line characteristics derived from fitted gaussians
are summarized in Table 1.

\section{HST Spectroscopy}

UV spectra of OI 363 and PKS1127$-$145 were obtained with the $HST$
Faint Object Spectrograph through the G160L/BL grating-digicon
combination. These QSOs were part of a large survey designed to
identify the nature of the Ly$\alpha$ line in known MgII
absorption-line systems. The survey spectra and results will be
presented elsewhere (Rao \& Turnshek, 1998).
 
The spectra of OI 363 and PKS1127$-$145 revealed DLa absorption lines
at the redshift of the intervening MgII absorption systems.  The
column density that best fit the data was determined by minimizing the
least squares difference between the data and the fit within the core
of the damped line.  The largest uncertainty in the column density
comes from continuum placement. Therefore, a conservative estimate of
the uncertainty was detemined by placing new continua at levels
corresponding to $1\sigma$ above and below the original continuum,
re-normalising the spectrum, and re-determining the best fit to the
damping profile in each case.  The column density of the OI 363
absorption-line system was measured to be 7.9$\pm$1.4 $\times 10^{20}$
cm$^{-2}$, while that of the PKS1127$-$145 system was measured to be
$5.1\pm0.9 \times 10^{21}$ cm$^{-2}$.

\section{Optical Data}

Optical observations of PKS B1127-145 and objects in the surrounding
field were made on 21 May 1990 with the EFOSC 2 instrument installed
at one of the Nasmyth foci of the ESO-3.5m NTT telescope, at La Silla,
Chile.  The detector used was the 1024 x 1024 pixels Thomson CCD \#
17, providing a pixel scale of 0.153 arcsec/pixel.  The observing run
was originally aimed at finding secondary images of highly-luminous
quasars due to gravitational lensing (cf. Surdej, et al., 1989). R-
and B-Band images with 5 minute exposures indicated the presence of
two faint companions to the QSO, with B-R colors compatible with those
expected for lensed secondary images. No standard stars were observed
in these filters.

In order to check if one or both companions were indeed a secondary
lensed image, two 30-minute spectra were obtained with grism \# 6 and
a 1.472$\arcsec$-wide slit oriented to include both objects and the QSO
(P.A. = 8.4$\deg$). The spectra were wavelength calibrated with He and
Ar arc lamp exposures taken immediately after the observation. The
wide slit, chosen to include as much light as possible from the
companion objects, leads to a poor resolution of only 27 \AA (FWHM).

The spectra of both companions indicate that they are galaxies, and
not additional, lensed images of the quasar.  Thus there is no
evidence for lensing in this system.  The companion object to the east
shows no emission lines, and the S/N in the continuum is not good
enough to determine its redshift.  However, the companion to the west
does show [OIII]4958 and [OII]5007 emission at $z = 0.3121 \pm
0.0003$, making it a candidate object for causing the neutral hydrogen
absorption at similar redshift.  The emission spectrum is shown in
fig. 3.

\section{Spin Temperature}

If a system is both a DLa absorber and an HI 21cm absorber, a
comparison of the neutral HI column density derived from the two
different absorption lines allows a determination of the mean spin
temperature, T$_s$, of the absorbing gas.  The equation relating
neutral hydrogen column density, N(HI), to the observed HI 21cm
absorption profile is: $$N(HI) =
1.8\times10^{18}~{T_s\over{f}}~EW_{21} ~{\rm cm}^{-2} $$ where f is
the fraction of the continuum source covered by the absorber, T$_s$ is
the spin temperature of the gas, and $EW_{21}$ is the integral of the
optical depth over velocity.  For a single line with a Gaussian
profile, $\tau(v)$, $$EW_{21} = 1.06\times\tau_{c}~\Delta V $$ where
$\tau_{c}$ is the peak optical depth of the line at the line center
and $\Delta V$ is the FWHM velocity in km s$^{-1}$.

Since both of the background radio quasars in this study are quite compact,
we assume that the covering factor, $f = 1$.  We also assume the same
gas is responsible for the 21cm absorption and the DLa, and use the
$N(HI)$ from the DLa line, to calculate the spin temperature of the
gas.  We find T$_s = 1230 \pm 335$ K for OI 363, and T$_s = 1000 \pm
200$ K for PKS B1127-145.

When calculated in this way, the spin temperature represents a column
density weighted harmonic mean temperature of all the gas along the
line of sight.  Since the gas is most likely composed of more than one
temperature phase, the spin temperature value does not have a
straightforward interpretation (cf. Carilli, et al. 1996).  However,
in the simplest sense, it can be understood as an upper limit to the
temperature of the gas in the coldest phase.

\section{Discussion}

\subsection{OI 363}
OI 363 (0738+313), is a core dominated quasar at $z_{em} = 0.630$.
Observations at 1640 MHz (Murphy, et al., 1993) show that the lobes extending
$\approx 30\arcsec$ from the core contain only $3\%$ of the total flux
of the quasar.  The quasar is slightly variable at low frequencies
(Bondi, et al., 1996b).

The metal line absorption system was originally reported by Boulade,
et al. (1987) at a redshift of $z = 0.2213$, and subsequently by
Boiss\'e, et al. (1992) at a redshift of $z = 0.2216 \pm 0.0003$.
The only identified lines in the spectrum are the MgII
$\lambda\lambda2796,2803$ doublet and a possible MgI line.  Deep
optical imaging was made by LeBrun et al. (1993), who identified
what they considered a likely absorber at a projected separation from
the quasar of 5.70'', or $1.24R_H$ if it lies at $z = 0.221$.  They
identified three additional galaxies, at smaller angular separations
from the quasar, which are very faint and would be dwarf galaxies at
the absorption redshift.  Unfortunately, they do not report a
confirmed redshift for any object in the field near the quasar.

The HI 21cm absorption line, shown in fig. 1, has a narrow width of
only two channels over most of its depth, and may be unresolved by
this observation.  This implies an upper limit to the line width of
$\sim8$ km s$^{-1}$ at the 6 km s$^{-1}$ resolution of the data.  The
HI column density from the DLa profile is $N(HI) = 7.9\pm1.4 \times
10^{20}$ cm$^{-2}$, and the calculated mean harmonic spin temperature
is T$_s = 1230 \pm 335$ K.  The termal kinetic temperature of the gas
for a line of width 8 km s$^{-1}$ is T$_k = 1400$ K.  This means that within
the errors, T$_s = $T$_k$.

\subsection{PKS B1127-145}

PKS B1127-145 is a compact, gigahertz peaked radio source at $z =
1.187$.  VLBI observations at 1670 MHz show a slightly elongated
structure with an extent of approximately 20 mas. Observations at 408
MHz give a flux variability of 1.2 Jy/year (Bondi, et al., 1996a), and
indicate structural variations as well.

Fig. 2 shows the neutral HI 21cm absorption for this system.  The
observed flux of the quasar is 5.25 Jy in this observation.  This is
somewhat lower than reported fluxes in NED\footnote{The NASA/IPAC
  Extragalactic Database (NED) is operated by the Jet Propulsion
  Laboratory, California Institute of Technology, under contract with
  the National Aeronautics and Space Administration.}, however, as
discussed above the quasar is a known variable at low frequency.  If
the absorption line is fit by one gaussian component, the optical
depth of the absorption is $6.2\%$, and the FWHM is approximately 42
km s$^{-1}$.  However, there is some evidence for structure in the
line.  In particular, the split in the middle appears to be real,
suggesting at least two components, and the asymmetric low frequency
side of the profile may result from a third component.  There is some
low level interference in the spectrum adjacent to the low frequency
side of the absorption line, at $\sim$1081.75 MHz, which makes
interpretation of the real shape difficult.  The HI column density
from the DLa profile is $N(HI) = 5.1\pm0.9 \times 10^{21}$ cm$^{-2}$,
and the calculated T$_s = 1000 \pm 200$ K.

Bergeron and Boiss\'e (1991) studied this $z_{abs}=0.313$ metal-line
system in some detail.  They identify MgII, FeII and MgI absorption in
the spectrum.  The average redshift of the metal lines is $z = 0.3127
\pm 0.0002$.  In addition to the absorption spectrum, and deep images
of the field, they present emission spectra for two of the bigger
bright galaxies near the quasar (different from the two faint
companions discussed in Sect. 4), both of which have redshifts similar
to that of the metal line system. They identify the galaxy at the
smaller projected distance from the quasar sightline as the absorber.

The ESO/EFOSC2 spectral observations (discussed in Sect. 3) identify a
new candidate absorber galaxy for this system. The new emission object
is a close companion to the west of the quasar and lies at a projected
distance of $3.9\arcsec$ or 11 $h^{-1}_{100}$kpc from the quasar
sightline.  Based on comparison with other objects in our image for
which apparent magnitudes are given in Bergeron and Boiss\'e (1991),
it has an apparent magnitude of $m_r = 22.3$.  The galaxy which
Bergeron and Boiss\'e identify as the absorber is at a projected
separation of $9.6\arcsec$, or 37 $h^{-1}_{100}$kpc from the quasar at
the redshift of the galaxy.  This corresponds to a galactic radius of
$2.7 R_H$.  A column density of neutral gas of $10^{21}$ cm$^{-2}$ is
unlikely at this galactic radius, and we consider the new emission
object, although smaller and fainter, to be the more likely absorber
given its proximity to the quasar sightline.  It is also possible that
the three galaxies at similar redshift have undergone strong
interaction with each other, in which case the absorbing gas could be
tidal debris.

\subsection{Spin Temperature}

Fig. 4 shows T$_s$ vs. redshift for all of the known HI 21cm and DLa
absorber systems, calculated from the literature (as summarized by
Carilli, et al., 1996), with the two new data points marked by open
symbols.  The shaded region shows the range of Galactic T$_s$ values
at optical depths comparable to those found in the DLa systems, using
numbers from Braun and Walterbos (1992).  The large errorbars in any
given measurement are dominated by uncertainties in the true optical
continuum level, due to confusion from the Ly$\alpha$ forest lines,
which make fitting the damped profile difficult.  As noted in previous
studies (cf. de Bruyn, et al.  1996), all of the redshifted absorbers
except the highest optical depth (lowest T$_s$) system have T$_s$
values which are roughly two or more times greater than the Galactic
values at similar optical depths.

Most estimates of T$_s$ for the present epoch have been based on
studies of the Milky Way or Andromeda (cf. Dickey and Brinks (1988);
Braun and Walterbos (1992), and references therein).  These estimates
use column densities derived from HI 21cm emission lines rather than
from DLa absorption lines, but studies have shown that
N(HI)$_{Ly\alpha} \approx$ N(HI)$_{21cm}$ (Dickey and Lockman, 1990) .
This suggests that T$_s$ values for the galaxy and the redshifted DLa
systems, although calculated from different quantities, can be
compared.

On the other hand, values for T$_s$ have a strong correlation with the
column density, N(HI), in each of the clouds along the line of sight
in the Galaxy, and are thus sensitive to the location of individual
gas clouds.  A single cloud usually has a greater angular size on the
sky than the beam with which it is observed, so it is possible to
calculate the spin temperature for just one cloud.  This is not the
case in redshifted systems, where a line of sight includes more of the
galaxy and possibly many clouds.  Thus it is difficult to usefully
compare present epoch spin temperature values to those calculated at
higher redshifts.

The new systems in this study were observed at low redshift with the
same line of sight limitations as the higher redshift systems,
allowing a meaningful comparison.  There is no clear trend for a
change in T$_s$ with increasing redshift among the eight systems shown
in Fig.  4.  If the gap between T$_s$ values in the galaxy and those
in the DLa systems is an effect of evolution over time, then all of
that evolution must have occured between $z = 0.2$ and the present
epoch.  Instead, we consider it more likely that the gap arises from
the presence of many clouds in one line of sight at high redshift, or
from differences in the radio and optical sightlines.

The sizes of the radio emission regions of quasars are much larger
than the optical regions, and usually larger than an average cloud as
well. It is therefore likely that the optical and radio lines of sight
actually sense different clouds in a redshifted galaxy, and hence
have different column densities of neutral gas.  This implies that the
spin temperatures derived by assuming both column densities are equal
may be meaningless.  Unfortunately, the resolution obtained in most
radio survey observations (with single dishes or synthesized beams
from arrays like the WSRT) is at least as large on the sky as compact
background radio sources, and gives no spatial information about the
clouds in front of the quasar.  Future use of VLBI techniques to
pinpoint the radio sightlines more accurately may help to clarify this
problem.

\begin{acknowledgements} 
  
  This research has made use of the NASA/IPAC Extragalactic Database
  (NED) which is operated by the Jet Propulsion Laboratory, California
  Institute of Technology, under contract with the National
  Aeronautics and Space Administration.

\end{acknowledgements}

\newpage 

\figcaption[lane.fig1.eps]{Spectrum from a 90 minute WSRT
    observation of OI 363 showing HI 21cm absorption at $z = 0.2212$.
    The rms noise is 14 mJy/channel, and the velocity spacing is 5 km
    s$^{-1}$.  Data have been given a natural weight and hanning
    smoothed but not resampled.  The expected frequency for absorption
    is based on a low resolution redshift for MgII absorption in this
    system (Boisse, et al.  1992).  The MgII absorption is much
    broader in velocity and comes from gas of lower HI column density,
    which accounts for its slight redshift offset from the HI 21cm
    line. \label{Figure 1}}

\figcaption[lane.fig2.eps]{Spectrum from a 60 minute WSRT observation
    of PKS B1127-145 showing HI 21cm absorption at $z = 0.3127$.  The
    rms noise is 14 mJy/channel, and the velocity spacing is 5.4 km
    s$^{-1}$.  Data have been given a natural weight and hanning
    smoothed, but not resampled.  The frequency at which absorption
    was expected based on the metal absorption lines (Bergeron and
    Boiss\'e, 1991) is marked and the standard deviation of this value
    is shown. \label{Figure 2}}

\figcaption[lane.fig3.ps]{ESO/EFOSC2 emission spectrum for small companion
  galaxy close to the line of sight PKS B1127-145, showing [OIII]4958
  and [OII]5007 emission at $z = 0.3121 \pm 0.0003$. \label{Figure 3}}

\figcaption[lane.fig4.eps]{Plot of T$_s$ vs. redshift, for all known high
    redshift HI 21cm/damped Ly$\alpha$ absorbers.  Open points from
    this study.  Filled squares calculated from the literature, as
    summarized by Carilli et al. (1996).  Values of optical depth for
    the absorbers fall in the range $0.02 < \tau < 0.7 $.  The shaded
    region shows the range of T$_s$ values in the Milky Way for
    optical depths in the range $0.01 < \tau < 1$ (Braun and
    Walterbos, 1992).  T$_s$ is generally larger for smaller optical
    depth systems. \label{Figure 4}}

\vskip 2.0cm
\begin{table}[h]
\caption[]{Absorption information}
\begin{flushleft}
\begin{tabular}{lll}
  \tableline 
  \tableline 
  ~~ & PKS B1127-145 & OI 363 \\ 
  \tableline 
  $z_{em}$ & 1.187 & 0.630 \\ 
  $z_{abs}$ & 0.3127 $\pm$ 0.0002 & 0.2212 $\pm$ 0.0001 \\ 
  21cm line depth (mJy) & 330 $\pm$ 14 & 82 $\pm$ 14 \\ 
  $S_\nu$ continuum (mJy) & 5285 $\pm$ 11 & 1940 $\pm$ 5 \\ 
  21cm optical depth, $\tau_{\rm c}$ & 0.062 $\pm$ 0.003 & 0.042 $\pm$ 0.007 \\
  21cm FWHM (km s$^{-1}$) & 42.1 $\pm$ 2.7 & 8 $\pm$ 1 \\ 
  N(HI)$_{DLa}$ (cm$^{-2}$) & $5.1\pm0.9 \times 10^{21}$ &
  $7.9\pm1.4 \times 10^{20}$ \\ 
  T$_s$ (K) & 1000 $\pm$ 200 & 1230 $\pm$ 335 \\ 
  \tableline
\end{tabular}
\end{flushleft}
\end{table}

\newpage

\begin{figure}[ht]
\centering
\centerline{Figure 1}
\leavevmode
\epsfxsize=.455\columnwidth
\epsfbox{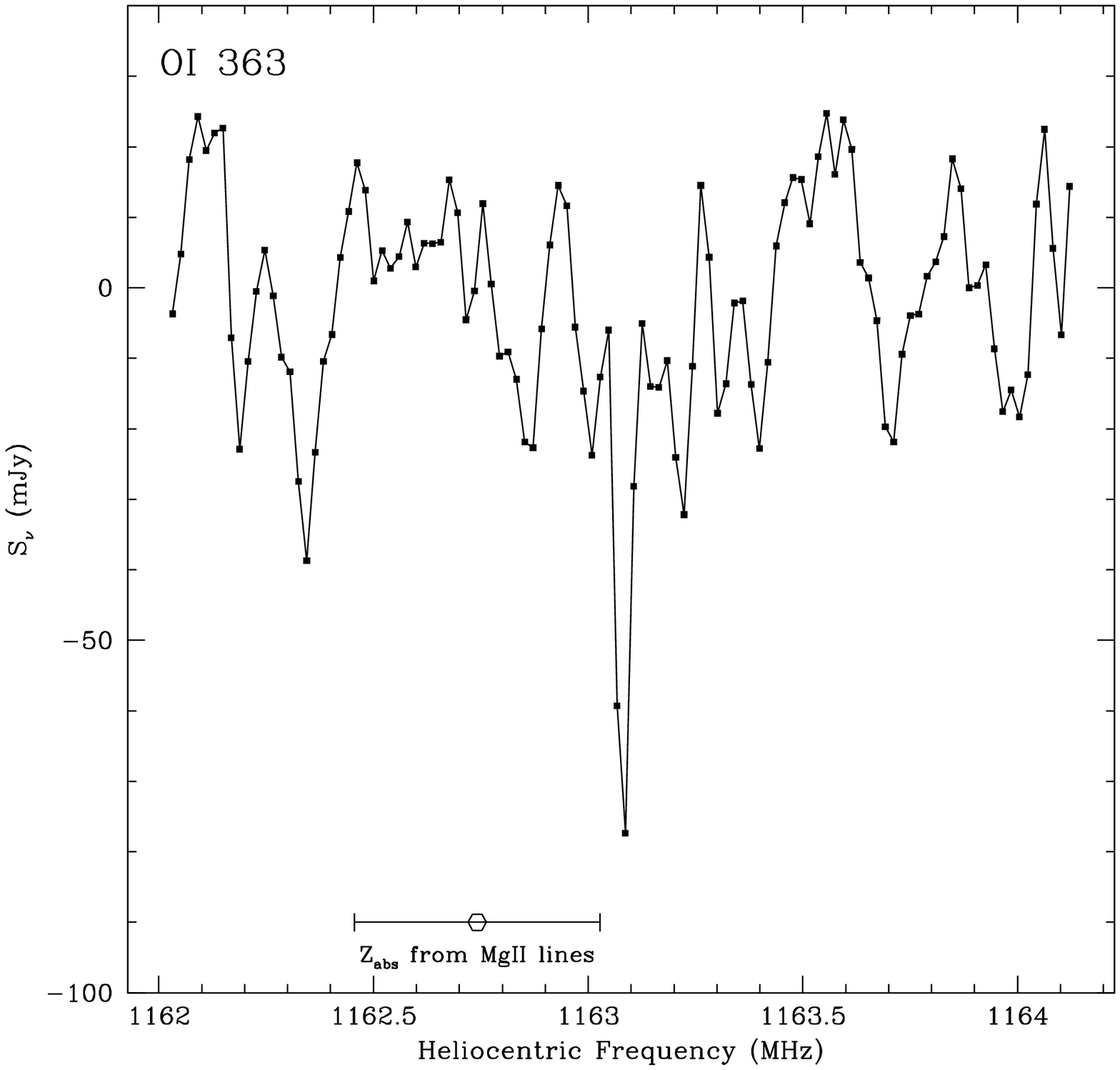}
\end{figure}

\vskip2.0cm

\begin{figure}[h]
\centering
\centerline{Figure 2}
\leavevmode
\epsfxsize=.455\columnwidth
\epsfbox{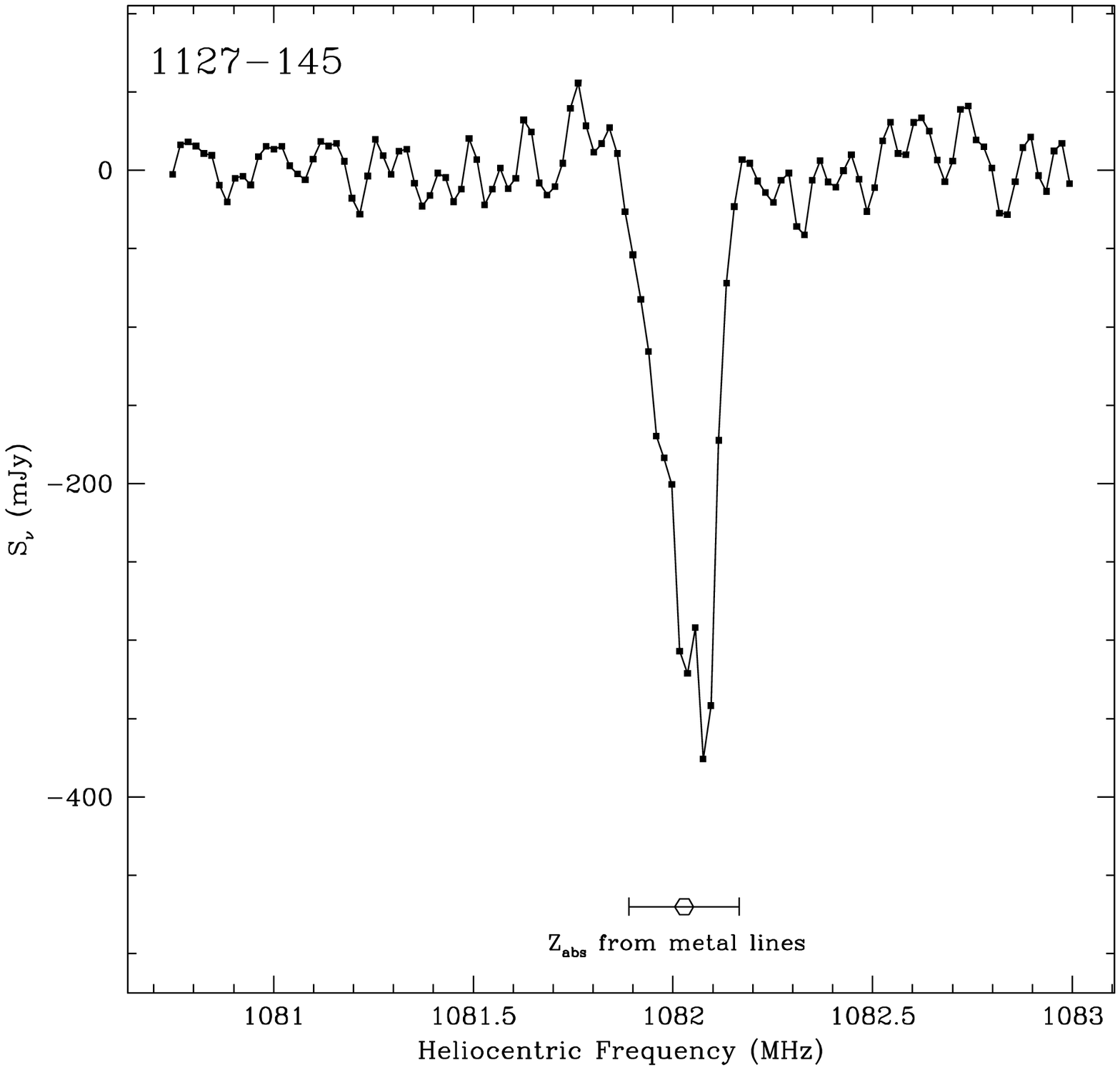}
\end{figure}

\newpage

\begin{figure}[ht]
\centering
\centerline{Figure 3}
\leavevmode\kern-15cm
\includegraphics{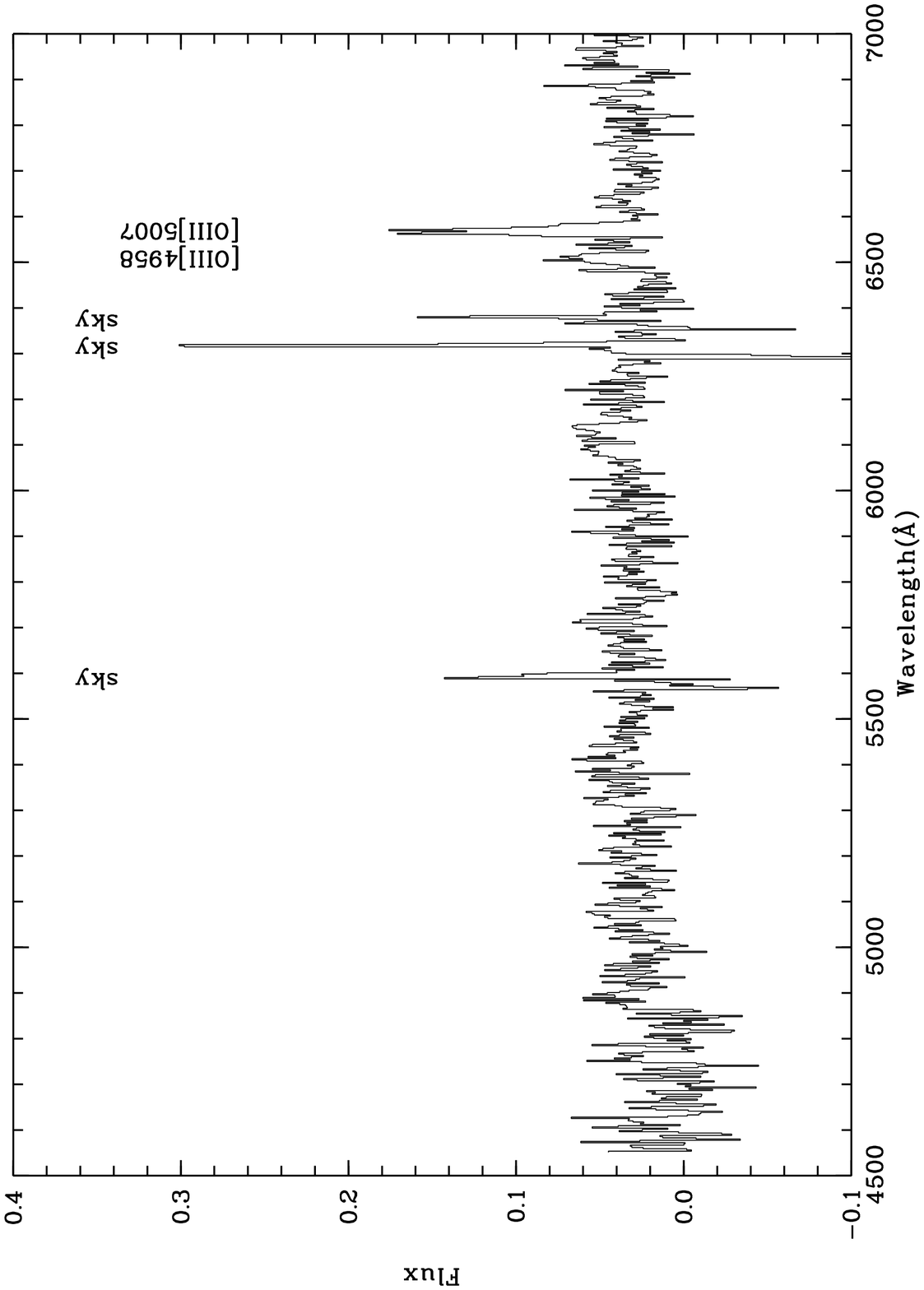}
\end{figure}

\null\vskip10cm
\begin{figure}[ht]
\centering
\centerline{Figure 4}
\leavevmode
\epsfxsize=.55\columnwidth
\epsfbox{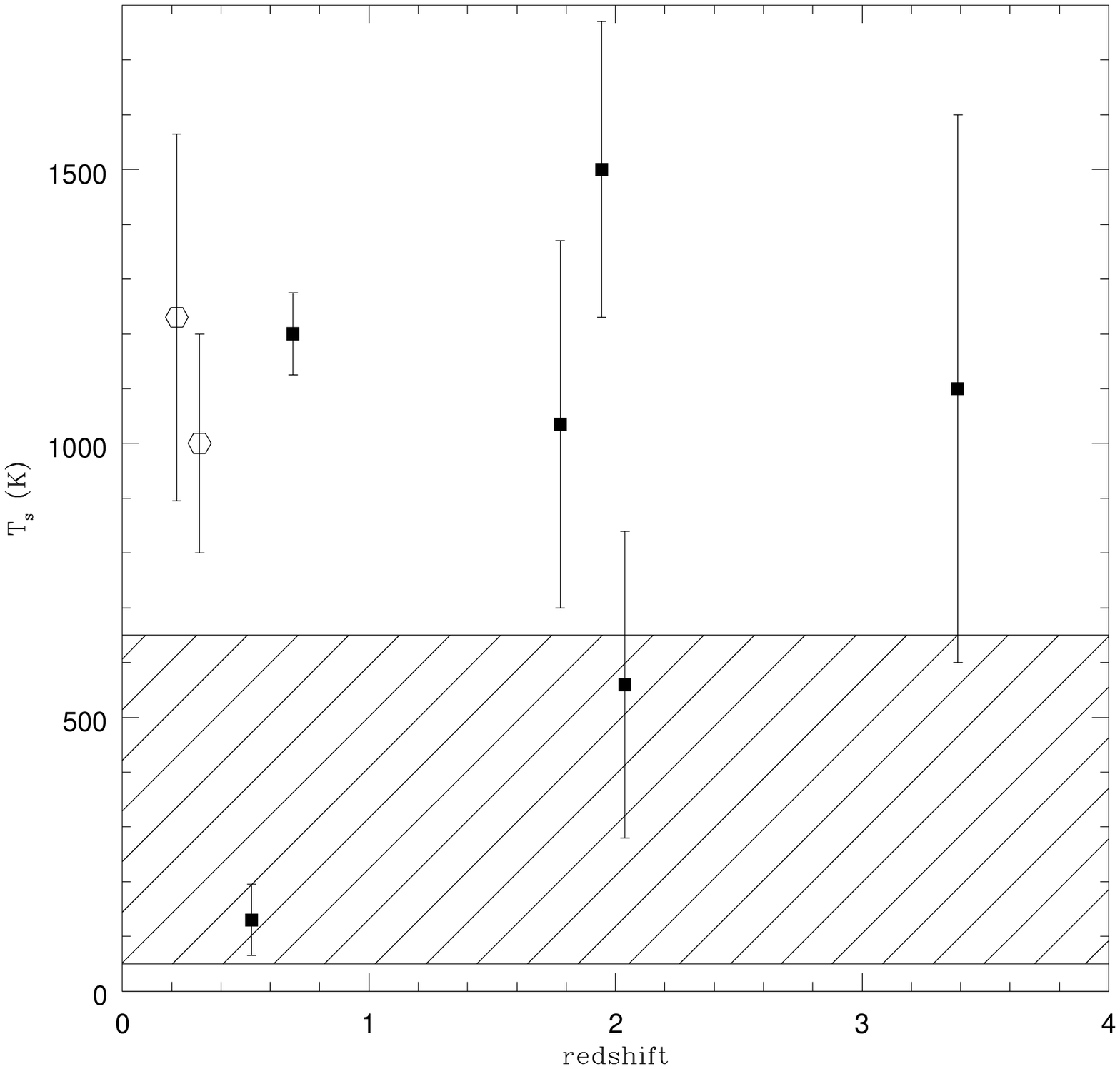}
\end{figure}

\end{document}